\documentclass[twocolumn,showpacs,floats,superscriptaddress,aps,pra]{revtex4-1}
\usepackage{amsfonts}
\usepackage{amssymb}
\usepackage{amsmath}
\usepackage{graphicx}
\usepackage{bm}
\usepackage{color}
\usepackage{float}

\usepackage{calc}
\usepackage{ifthen}
\usepackage{mwe}
\usepackage{subfigure}
\DeclareUnicodeCharacter{2212}{:}
\usepackage{natbib}
\usepackage{epstopdf}
\usepackage{epsfig}
\usepackage{braket}
\usepackage{hyperref} 
\hypersetup{
    colorlinks = true,
    linkcolor = {blue},
    citecolor = {blue},
}

\begin{document}
\raggedbottom

\title{Detuning-modulated universal composite pulses}

\date{\today}

\author{Hadar Greener}
\thanks{These authors contributed equally}
\affiliation{Condensed Matter Physics Department, School of Physics and Astronomy, Tel Aviv University, Tel Aviv 69978, Israel}
\affiliation{Center for Light-Matter Interaction, Tel Aviv University, Tel Aviv 69978, Israel}

\author{Elica Kyoseva}
\thanks{These authors contributed equally}
\email{Current address: Boehringer Ingelheim, Tel Aviv Israel}
\affiliation{Condensed Matter Physics Department, School of Physics and Astronomy, Tel Aviv University, Tel Aviv 69978, Israel}
\affiliation{Center for Light-Matter Interaction, Tel Aviv University, Tel Aviv 69978, Israel}

\author{Haim Suchowski}
\email{haimsu@tauex.tau.ac.il}
\affiliation{Condensed Matter Physics Department, School of Physics and Astronomy, Tel Aviv University, Tel Aviv 69978, Israel}
\affiliation{Center for Light-Matter Interaction, Tel Aviv University, Tel Aviv 69978, Israel}

\begin{abstract}
We present a general method to derive detuning-modualted composite pulses (DMCPs) as N rotations of a canonical two-state quantum system to create accurate and robust pulses that are independent of the initial state of the system. This scheme has minimal pulse overhead, and achieves pulses that are stable against amplitude errors well within the $10^{-4}$ threshold that may be suitable for quantum information processing (QIP), within the lifetime of the system. This family of pulses enables to overcome inevitable fabrication errors in silicon photonics, and relax the need for a precise initial state of light coupled into the system to achieve accurate light transfer. Furthermore, we extend universal DMCPs to n-level systems with irreducible SU(2) symmetry to create state transfer that is highly robust to errors in the pulse area from any initial state.   
\end{abstract}

\maketitle
\section{Introduction \label{sec:1}}
Recent advances in quantum computation provide an actionable platform for scaling up the number of qubits in a circuit in the foreseeable future. Any quantum circuit hardware performance relies on high-fidelity unitary quantum gates, allowing for precise state preparation and transfer. These gates are the fundamental building blocks of quantum information processing (QIP), where the admissible error of quantum operations \cite{QuantumInfo1} is smaller than $10^{-4}$. This is a challenging limit  in experimental realizations of QIP, where any systematic error can reduce the state fidelity below this fault-tolerance. The slightest fabrication defects or an inaccurate coupling strength can lead to such errors, that include deviations from target driving amplitudes and frequencies. This is approached by various methods, including closed-form solutions based on the mathematical group theory description of a quantum system, commonly referred to as the field of geometric quantum coherent control. Examples of analytic time dependent models that were derived for precise state transfer include adiabatic evolution schemes \cite{Zener,RosenZener,AllenEberly,STIRAP} that utilize a very slow change of one of the control parameters compared to another in order to reach the target state with high precision. While this family of solutions is very popular, these are asymptotic solutions that require long control sequences as well as the preservation of the adiabatic criteria during the state evolution.

Another method to correct for such errors are composite pulses (CPs) \cite{CP1,CarrSpinEcho,CP2,CP3,CP4,CP5,CP6,CP7}. These are a series of pulses with specifically calculated areas and phases that, when applied in sequence, create accurate and robust quantum gates. CPs were historically designed for resonant or adiabatic interactions with complex coupling parameters \cite{VitanovSmooth,GenovPRL}, realizing complete population transfer (CPT) in physical systems by radiofrequency (rf) and ultrashort pulsed excitations \cite{KeelerNMR}. More recently, CPs have been applied for quantum information processing \cite{jones2009composite,husain2013further,alexander2020qiskit}. However, CPs were unable to offer robust solutions for applications where the coupling is not complex valued, can not be controlled or does not exist. One particular instance for this is integrated photonic circuits, where the coupling parameters are always real-valued. This is an important limitation to address in CPs, since photonic systems are a promising contender for quantum computation hardware, due to on-chip integration capacity and scalability. Despite this, their immediate implementation is hindered by inevitable fabrication errors that result in gate operations that do not meet the QIP standards.

Recently developed detuning-modulated composite pulse (DMCP) sequences \cite{PhysRevA.100.032333} address this exact issue of systems with real coupling parameters and off-resonance control, and have been utilized for state-to-state transfer in any qubit architecture, including integrated photonic circuits. For this particular realization, the application of DMCPs is crucial for precise photonic-based qubit gates. In order to implement accurate quantum algorithms, a robust realization of unitary gates that are independent of the initial state of the qubit should be readily available. However, to date, DMCP sequences were derived only as a robust mechanism for state-to-state transfer, mainly from the ground state.

Here we present the first universal DMCPs for the implementation of robust state transfer from any initial state, excluding any constraint on the coupling strength. Our sequences exhibit high fidelity in the presence of errors in various systematic parameters including pulse area (under $10^{-4}$), coupling and detuning (under $10^{-2}$) - all well within the qubit’s lifetime. We provide a general method to derive universal DMCPs as $N$ rotations of a canonical two-state quantum system and generalize to higher-level quantum systems with irreducible SU(2) symmetry. This scheme has a minimal pulse overhead, such that robust realizations of universal DMCPs are as short as $N=4$. However, while these pulses were designed to achieve stability to amplitude errors of the unitary propagator, its phase is not. Further derivations of unitary gates that are stable against amplitude and phase errors are beyond the scope of this work.

\begin{figure}[tb]
\centering
\includegraphics[scale=0.25]{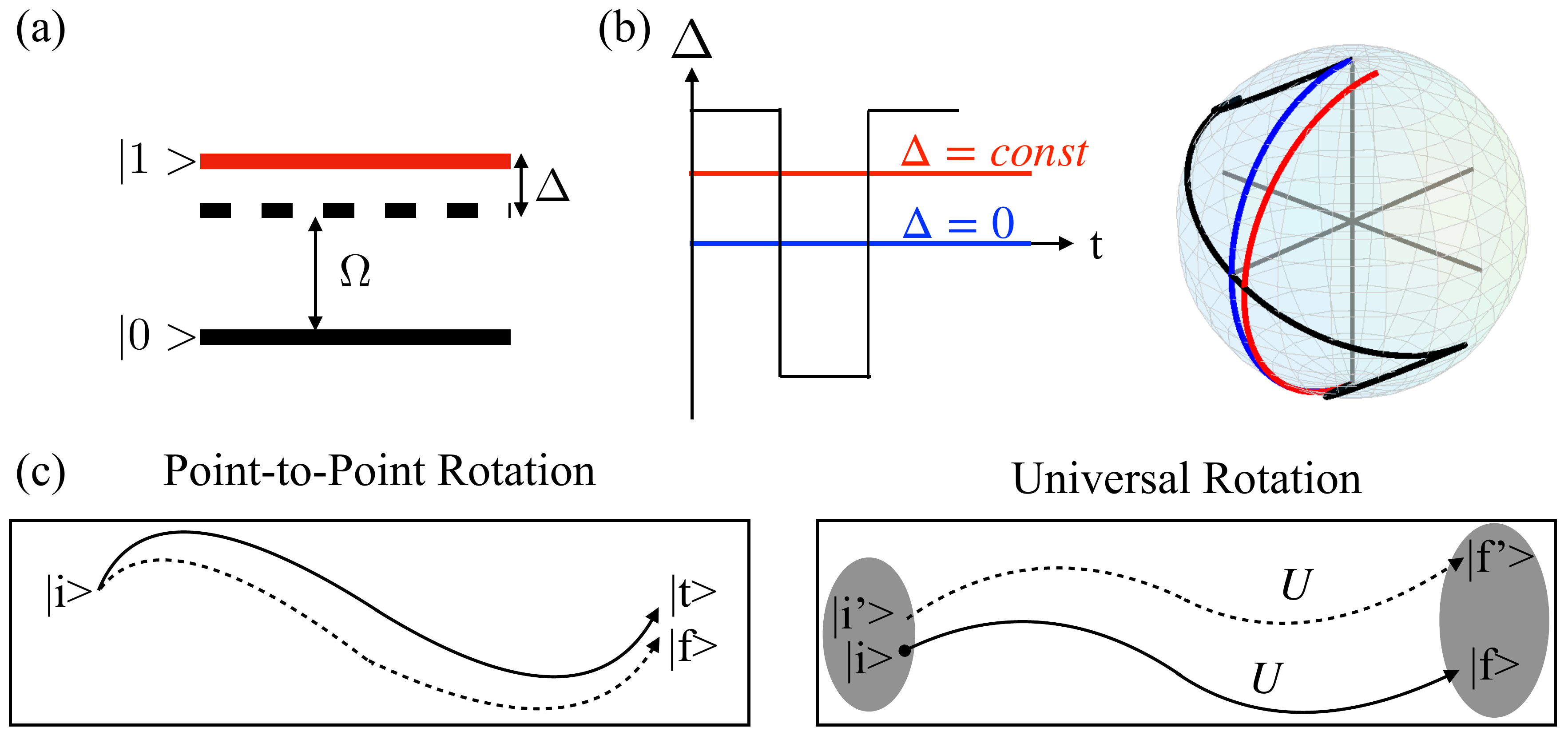}
\caption{\textbf{Universal detuning-modulated composite pulses.} (a) Schematic depiction of a qubit, a two-level quantum system with coupling $\Omega$ and detuning $\Delta$. (b) Detuning as a function of time for a resonant (blue), constant detuned (red) and detuning-modulated (black) system with their respective trajectories on the Bloch sphere. (c) State-to-state rotation (left) from initial state $|i>$ to desired target state $|t>$ (continuous line) vs. erroneous experimental rotation to final state $|f>$. Universal rotations (right) $U$ from initial state $|i>$ to target state $|f>$ preserves the rotation from \emph{any} other initial state $|i'>$ to the final rotated state $|f'>$.
\label{Figure:Fig1}}
\end{figure}

%%%%%%%%%%%%%%%%%%%%%%%%%%%%%%%%%%%%%%%%%%%%%%%
%%%%%%%%%%%%%%%%%%%%%%%%%%%%%%%%%%%%%%%%%%%%%%%
%%%%%%%%%%%%%%%%%%%%%%%%%%%%%%%%%%%%%%%%%%%%%%%
\section{Universal Detuning Modulated Composite Pulses \label{sec:2}}
%%%%%%%%%%%%%%%%%%%%%%%%%%%%%%%%%%%%%%%%%%%%%%%
%%%%%%%%%%%%%%%%%%%%%%%%%%%%%%%%%%%%%%%%%%%%%%%
%%%%%%%%%%%%%%%%%%%%%%%%%%%%%%%%%%%%%%%%%%%%%%%
\begin{figure}[tb]
\centering
\includegraphics[scale=0.25]{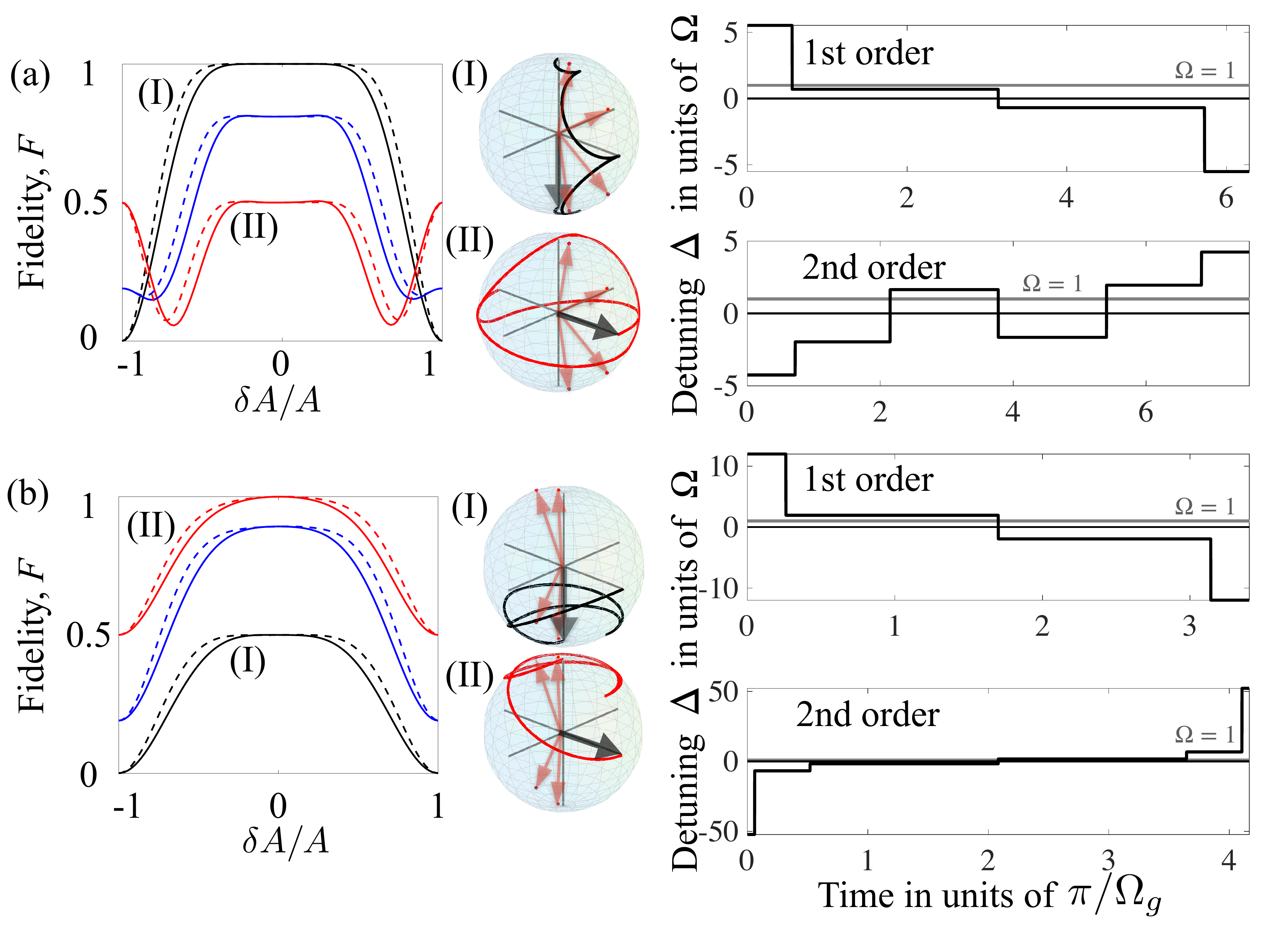}
\caption{\textbf{Fidelity of detuning-modulated universal composite pulses.} (left) First-(continuous lines) and second-order (dashed lines) (a) $\pi$ and (b) $\pi/2$ pulses vs. errors in the pulse area for different initial states $|0\rangle$ (I - black), $1/\sqrt{2}(|0 \rangle +|1\rangle)$ (II - red) and $0.9(|0 \rangle +\sqrt{0.19}|1\rangle)$ (blue) with matching (center) first-order trajectories on the Bloch sphere. These trajectories are plotted around the  vector frequencies $-\Omega_{i}/\Omega_{g,i}\hat{x}+\Delta_{i}/\Omega_{g,i}\hat{z}$ (red arrows), which directly correspond to the axes offsets of the propagator $V(v)$ (see text for details). The construction of the universal pulses by detuning modulation (right) for first- (N=4) and second-order (N=6) pulse sequences as a function of time in units of the generalized Rabi frequency. All sequences are demonstrated with constant and normalized coupling $\Omega = 1$ (see text for details and definition of fidelity).
\label{Figure:Figure2}}
\end{figure}

We begin to lay out the basis for universal detuning-modulated composite pulses by differentiating between two different types of pulses: point-to-point (PP) and universal rotations (UR). The first are constructed to transform a given state to a desired final state, while the latter is designed to create a given rotation around a specific axis and angle for any arbitrary initial state. The PP solutions based on DMCPs have been developed recently \cite{PhysRevA.100.032333} for the most general qubit system.

The Schr\"{o}dinger equation governs the unitary evolution of a qubit system \{$|0\rangle, |1\rangle$\} shown in Fig. \ref{Figure:Fig1} driven coherently by an external electromagnetic field in the following way: 

\begin{equation}
\centering
i \hbar \partial_t \left[ \begin{array}{c}
c_0(t) \\
c_1(t) \end{array} \right] = \frac{\hbar}{2} 
\left[ \begin{array}{cc}
-\Delta(t) & \Omega(t) \\
\Omega^*(t) & \Delta(t) \end{array} \right] 
\left[ \begin{array}{c}
c_0(t) \\
c_1(t) \end{array} \right] ,
\label{Sch}
\end{equation}

where $[c_0(t),c_1(t)]^{\textnormal{T}}$ is the probability amplitudes vector, $\Omega(t)$ is the coupling between the states of the system, and $\Delta(t) = \omega_0 - \omega$ is the real-valued detuning between the excitation frequency $\omega$ and the qubit frequency $\omega_0$. In the following, we assume $\Omega(t)$ and $\Delta(t)$ real and constant for each constituent pulse in the sequence; however, the proposed derivation is straightforward to extend to complex values.

The unitary propagator of the time evolution governed by Eq. \eqref{Sch} is found according to $U(t,0) = e^{-i/\hbar \int_{0}^{t} H(t) dt}$. In the case that the Hamiltonian is constant in the time interval, the unitary propagator can be written as follows:

\begin{equation}
\centering
U(\delta t) = \left[ \begin{array}{cc}
\cos \left(\frac{A}{2} \right) + i \frac{ \Delta}{\Omega_g} \sin \left(\frac{A}{2} \right) & -i \frac{ \Omega}{ \Omega_g } \sin \left(\frac{A}{2} \right) \\
- i\frac{\Omega}{ \Omega_g } \sin \left( \frac{A}{2} \right) & \cos \left(\frac{A}{2} \right) - i\frac{ \Delta}{\Omega_g} \sin \left( \frac{A}{2} \right) \end{array} \right],
\label{U}
\end{equation}

where $\Omega_g = \sqrt{ \Omega^2 + \Delta^2}$ is the generalized Rabi frequency and $A =  \Omega_g \delta t$ is the pulse area with $\delta t =(t-t_0) $ the pulse duration. The evolution of the state of the qubit by the propagator $U(\delta t)$ from the initial time $t_0$ to the final time $t$ is $\mathbf{c}(t) = U(\delta t) \mathbf{c}(t_0)$. 

Robust universal pulses are generally difficult to implement with tailored pulses. Previous works in composite pulses \cite{VitanovSmooth,Elica:passband,PhysRevA.100.032333} have focused on achieving a stable bit flip realization only for an initial state of the system of either $|0\rangle$ or $|1\rangle$. Below, we show the first universal  DMCPs for implementing single qubit state transfer for any initial qubit state in a robust manner. 
Our proposed method is based on the recent methodology reported for deriving detuning-modulated state-to-state CPs comprised of $N$ individual off-resonant pulses with Rabi frequencies $\Omega_n$ and detunings $\Delta_n$ \cite{PhysRevA.100.032333}. State-to-state CPs are very sensitive to the initial state of the given system. Given the individual pulse propagator $U_n(\delta t_n)$ from Eq. \eqref{U}, the propagator for the total composite pulse sequence is given by the product 
\begin{equation}
\centering
U^{(N)}(T,0) = U_{N}(\delta t_N)\;U_{N-1}(\delta t_{N-1}) \dots U_{1}(\delta t_1),
\label{U_tot}
\end{equation}
where $\delta t_n = (t_{n} - t_{n-1})$ is the duration of the $n^{\textnormal{th}}$ pulse ($t_0 = 0$ and $t_N \equiv T$). For the exact form of the propagator elements we refer the reader to Ref. \cite{PhysRevA.100.032333}. 

Now we create the most general DMCPs-based solutions for state-independent single qubit pulses that are robust to various system inaccuracies and conclude in high-fidelity states. Namely, in contrast to the state-to-state solutions, universal DMCPs are not sensitive to the initial state of the system. These pulses are achieved by applying a procedure adapted from a previous work \cite{LUY2005179} for constructing a universal rotation (UR) of an angle $\theta$ from a point-to-point (PP)rotation of half the angle $\theta/2$. 

In order to create a rotation by the angle $\theta$ around axis $k$, the unitary transformation $U_{k}(\theta) = e^{-i\theta I_{k}}$ can be decomposed to two consecutive rotations of angle $\theta/2$ around the $k$ axis. Moreover, it has been shown \cite{LUY2005179} that a $\theta$ rotation around the the $\hat{x}$ axis can be decomposed to two PP pulse sequences:

\begin{equation}
\centering
    U_{x}(\theta) = V(v)\overline{V}^{tr}(v),
    \label{Khaneja_eqn}
\end{equation}

where $V(v)$ is the propagator describing a $\theta/2$ rotation PP sequence around a given range of axis offsets $v$ and $\overline{V}^{tr}(v)$ is its time and phase-reversed counterpart. Namely, a UR pulse is constructed by concatenating the original pulse sequence $V(v)$ to the time and phase-reversed pulse sequence $\overline{V}^{tr}(v)$. The phase reversal can be done around any axis; for e.g. $\hat{z}$, this results in $-\hat{z}$, which is a sign reversal of the detuning parameter. Note that for $\hat{x}$, since $\phi = 0$, phase reversal of the pulse results in $\hat{x}$.

\begin{figure}[tb]
\centering
\includegraphics[scale=0.25]{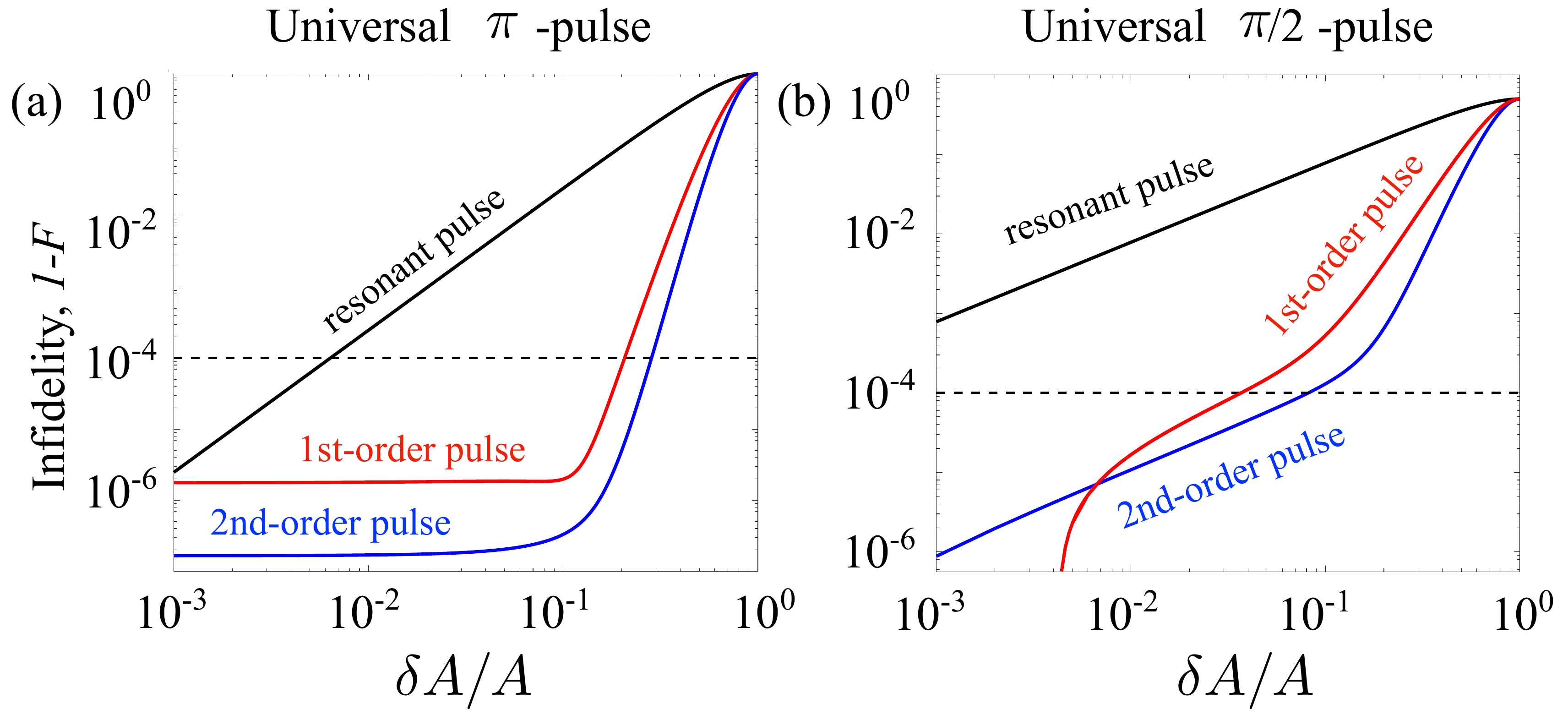}
\caption{\textbf{Infidelity (1-F) in logarithmic scale of detuning-modulated universal unitary single-qubit gates.} Infidelity of first-(red) and second-order (blue) (a) $\pi$ and (b) $\pi/2$ gates to pulse area errors. These pulses outperform single resonant pulse gates, plotted for reference in black. They maintain robustness within the $10^{-4}$ QIP infidelity threshold, shown as a dashed black line, for errors of up to at least $10\%$ from the target pulse area value.
\label{Figure:Figure3}}
\end{figure}

\begin{figure}[tb]
\centering
\includegraphics[scale=0.25]{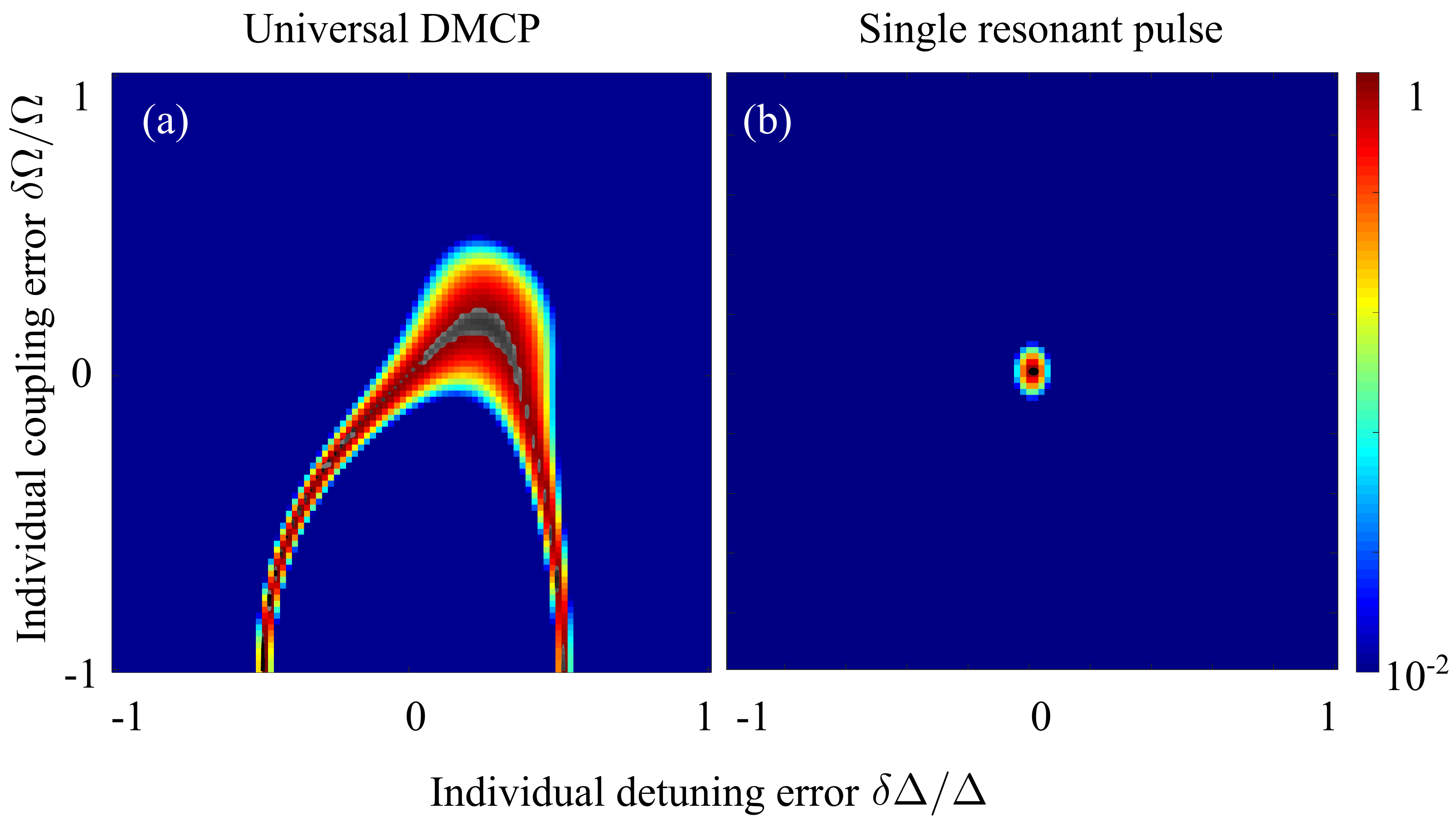}
\caption{\textbf{Contour plots of the robustness of universal unitary detuning-modulated pulses.} (a) Fidelity of a 1st-order universal unitary $\pi$ pulse against errors in individual detuning $\delta \Delta/\Delta$ and coupling $\delta \Omega / \Omega$ errors. The color scale ranges from a fidelity of $10^{-2}$ to a fidelity of $10^{0}$; in grey scale we accent the contour for which the fidelity is within the $10^{-4}$ QIP threshold. (b) Fidelity of a single resonant pulse against errors in the coupling and detuning. A single point on this contour plot achieves fidelity higher than $10^{-4}$.   
\label{Figure:Figure4}
}
\end{figure}

Focusing on a \emph{detuning}-modulated control sequence, we lay out the steps to construct a $N-$ piece universal detuning-modulated $\theta$ pulse.

\begin{enumerate}
    \item For $V(v)$, calculate a $N/2$-piece detuning-modulated $\theta/2$ pulse sequence with $N/2$ values of $\Delta = {\Delta_{1},...,\Delta_{N/2}}$, where the rotation axis offsets $v$ are calculated according to the resulting Bloch vector frequency $-\Omega_{i}/\Omega_{g,i}\hat{x} + \Delta_{i}/\Omega_{g,i}\hat{z}$ (see Fig. \ref{Figure:Figure2}).
    \item The detuning-reversed $\overline{V}^{tr}(v)$ is the above sequence with $\overline \Delta = {-\Delta_{N/2},...,-\Delta_{1}}$.
    \item The complete $N-$ piece universal detuning-modulated sequence is the concatenation of the two seqeuences.
\end{enumerate}

In the following, we devise the minimal \emph{first-order} universal detuning-modulated composite sequences. These are the family of solutions that nullify up to the fourth derivatives of the off diagonal element of the total propagator in Eq. \ref{U_tot} with respect to the pulse area around $A=\pi$ \cite{PhysRevA.100.032333}.

It has been shown \cite{PhysRevA.100.032333} that the shortest detuning-modulated pulses are composed of $N=2$ pieces. Thus, the shortest universal detuning-modulated $\pi$ pulse will be composed of $N=4$ pieces. In order to derive it, we calculated a $\pi/2$ detuning-modulated pulse sequence composed of $N/2=2$-pieces as $V(v)$ with two values of $\Delta = (\Delta_{1},\Delta_{2})$. An equal superposition between the two qubit states is realized when a single-qubit rotation T is at an angle $\theta = \pi/2$:

\begin{equation}
\centering
T = \left[ \begin{array}{cc}
\cos\theta & -i \sin \theta \\
-i \sin\theta & \cos\theta \end{array} \right].
\label{T}
\end{equation}

For $N/2=2$, the condition for the off-diagonal DMCP propagator element reads \cite{PhysRevA.100.032333}:
\begin{equation}
\centering
|U_{12}|^{2} = \frac{\frac{\Delta_1}{\Omega_1} - \frac{\Delta_2}{\Omega_2} }{\sqrt{1+\frac{\Delta_1}{\Omega_1}} \sqrt{1+\frac{\Delta_2}{\Omega_2}}} = - 1/\sqrt{2}. 
\end{equation}

Solving this equation for one of the independent parameters, $\frac{\Delta_1}{\Omega_1}$, we find that it is satisfied for $\frac{\Delta_1}{\Omega_1} = (-1 - \frac{\Delta_2}{\Omega_2})/(-1 + \frac{\Delta_2}{\Omega_2})$ and $\frac{\Delta_1}{\Omega_1} = (-1 + \frac{\Delta_2}{\Omega_2})(1 + \frac{\Delta_2}{\Omega_2})$. For a first-order sequence, this solution is plugged into the second derivative of $|U_{12}|^2$ with respect to $A$ at $A = \pi$ and we find its roots:

\begin{equation}
\centering
\Big( \frac{\Delta _ 1}{\Omega _1},  \frac{\Delta _ 2}{\Omega _2} \Big) = \pm (5.52, 0.69). 
\end{equation}

This gives the interaction parameters of a 2-piece sequence that produces a robust detuning-modulated $\pi/2$ pulse \cite{PhysRevA.100.032333}. Given this and the above formalism, the shortest universal DMCP is $\Delta_{i}=\pm (5.52, 0.69, -0.69,-5.52) \, \Omega$. This sequence enables a $\pi$ rotation from any initial qubit state, that is highly robust to pulse area errors.

To increase the robustness of our sequences to the above error, we solved for the family of \emph{second-order} universal detuning-modulated pulses, nullifying up to the sixth derivative of the modulus squared value of Eq. \ref{U_tot} with respect to $A$ at $A = \pi$. These result in sequences with a minimal length of $N=6$. We provide the shortest solutions for first- and second-order universal detuning-modulated $\pi$ pulses in Table 1.

\begin{table}
\centering
\begin{tabular}{ c c c c } 
 \hline \hline
 $N$ & Order & $(\frac{\Delta_{1}}{\Omega_{1}},\frac{\Delta_{2}}{\Omega_{2}}, \dots,-\frac{\Delta_{2}}{\Omega_{2}}, -\frac{\Delta_{1}}{\Omega_{1}})$\\ 
 \hline 
4 & 1 & $\pm (5.52, 0.69, -0.69,-5.52)$ \\ 
6 & 1 & $(5.89,1.01,-5.68,5.68,-1.01,-5.89)$ \\
6 & 2 & $(-4.25,-1.96,1.65,-1.65,1.96,4.25)$ \\ 
\hline
\end{tabular}
\caption{Detuning parameters for universal detuning-modulated composite $\pi$ pulses.}
\label{table:Table1}
\end{table}

In order to create universal and robust single-qubit $\pi/2$ pulses, two steps are required. The above technique is used to first derive a detuning-modulated $N/2=2$-piece $\pi/4$ pulse sequence. Then we reverse the sign of the detuning to create a universal sequence with $\Delta_{i}=(11.99,1.94,-1.94,-11.99) \, \Omega$. The shortest sequences for first- and second-order universal detuning-modulated composite $\pi/2$ pulses are shown in Table 2. 

\begin{table}
\centering
\begin{tabular}{ c c c c } 
 \hline \hline
 $N$ & Order & $(\frac{\Delta_{1}}{\Omega_{1}},\frac{\Delta_{2}}{\Omega_{2}}, \dots,-\frac{\Delta_{2}}{\Omega_{2}}, -\frac{\Delta_{1}}{\Omega_{1}})$ \\ 
 \hline 
4 & 1 & $(11.99,1.94,-1.94,-11.99)$ \\ 
6 & 1 & $(-0.97,0.97,0.37,-0.37,-0.97,0.97)$ \\
6 & 2 & $(-52.23,-6.76,-1.74,1.74,6.76,52.23)$ \\ 
\hline
\end{tabular}
\caption{Detuning parameters for universal detuning-modulated composite $\pi/2$ pulses.}
\label{table:Table2}
\end{table}

We tested the fidelity of universal DMCPs against various target system parameters. In this work, we define fidelity as:
\begin{equation}
    \centering
    F = |<\Psi_{realized}|\Psi_{target}>|^{2}
\end{equation}

We evaluate the robustness of the pulses as a function of target values of pulse areas for different initial states (see Fig. \ref{Figure:Figure2}). The robustness to errors maintains a high fidelity that is independent of the system's initial state, and increases for the second-order pulses plotted as dashed lines. We plot the infidelity of these pulses in logarithmic scale (see Fig. \ref{Figure:Figure3}) and compare it to that of a single resonant pulse and the QIP infidelity threshold of $10^{-4}$ for reference. The universal unitary $\pi$ pulses display robustness to errors of up to $28\%$ in the target pulse area while the universal $\pi/2$ pulses are robust to errors of up to $8\%$, compared e.g. with the $0.6\%$ robustness of a single resonant $\pi$ pulse.

In the detuning-modulated case, the pulse amplitude is a function of the coupling and the detuning. Any error in the pulse amplitude can be attributed to an error in the time in which the pulse was impinged on the system (as in the previous analysis) or to an error in either the coupling or the detuning. Thus, we studied the fidelity of the pulses as a function of the target detuning and coupling values, under the assumption that the realized time for each pulse was the target time (errorless). Fig. \ref{Figure:Figure4}(a) is a contour plot of the first-order universal detuning-modulated $\pi$ pulse fidelity, as a function of errors in the individual target coupling and detuning values. The color scale of this plot is from a fidelity of $10^{-2}$ to $10^{0}$, and we accent in greyscale the contour of the combined values of coupling and detuning that achieve fidelities above $10^{-4}$. For a single resonant pulse, this contour is minimized to a single black point of an errorless coupling value, as seen in Fig. \ref{Figure:Figure4}(b). We provide full-ranged contour plots for the fidelity of the universal DMCPs against errors in the target coupling and detuning values in Appendix A.

\begin{figure}[h!]
\centering
\includegraphics[scale=0.25]{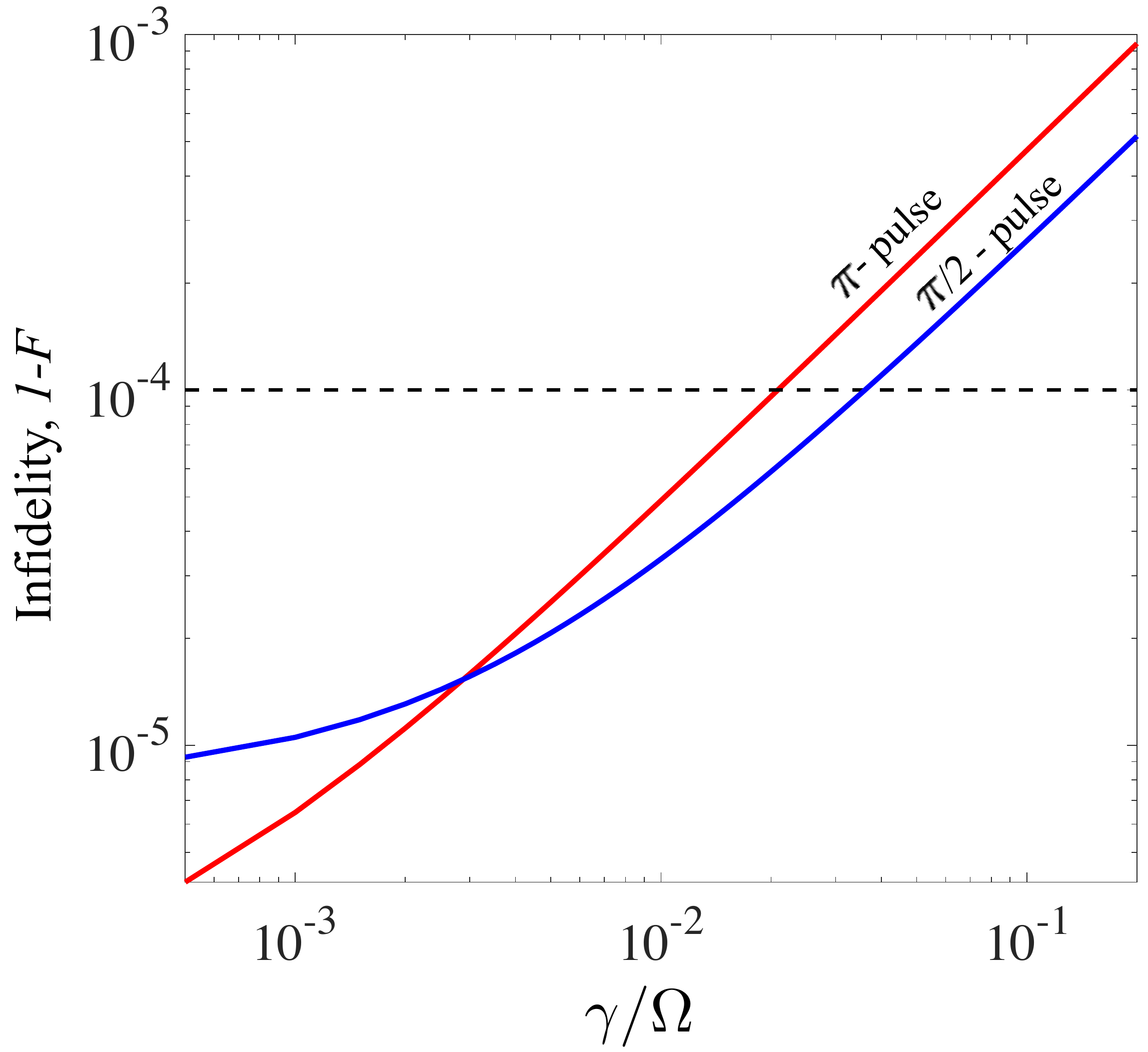}
\caption{\textbf{Robustness of universal detuning-modulated CPs vs relaxation.} Infidelity, $1-F$, of universal $\pi$ (red) and $\pi/2$ (blue) pulses in logarithmic scale as a function the system's decay rate in units of $\Omega$. The pulses maintain robustness within the $10^{−4}$ QIP infidelity threshold, shown as a dashed black line, for decay rates lower than $10\%$ of the system's coupling value. The infidelity of a single resonant pulse is out of the scale of this graph.
\label{Figure:Figure6}}
\end{figure}

As composite sequences are comprised of a series of pulses, their overall implementation time is longer than that of a single resonant pulse. Therefore, one must also test their fidelity against the system's lifetime. Substituting $\Delta \rightarrow \Delta - i\gamma$ in the diagonal elements of the Hamiltonian presented in the main text, we find the probability amplitude of each state according to $|c_{i}(t)|^{2}e^{-\gamma t/2}$, where $\gamma$ is the characteristic relaxation time of the system is $T_{1} = \gamma^{-1}$. For free decay, $T_{1}$ is independent of $T_{2}$, and there is an upper limit \cite{relaxation1,relaxation2} for the decoherence rate $T_{2} \leq 2T_{1}$. Fig. \ref{Figure:Figure6} presents the robustness of both the $\pi$ and $\pi/2$ pulses with respect to $\gamma$. We considered experimentally reported values of $\gamma$ of the order of $\Omega$ \cite{PhysRevLett.93.157005,Decoherence1,PhysRevB.74.161203,Decoherence2} to show that our pulses are robust to decoherence and that their implementation time is well within the decay rate of the qubit.

\begin{figure}[tb]
\centering
\includegraphics[scale=0.23]{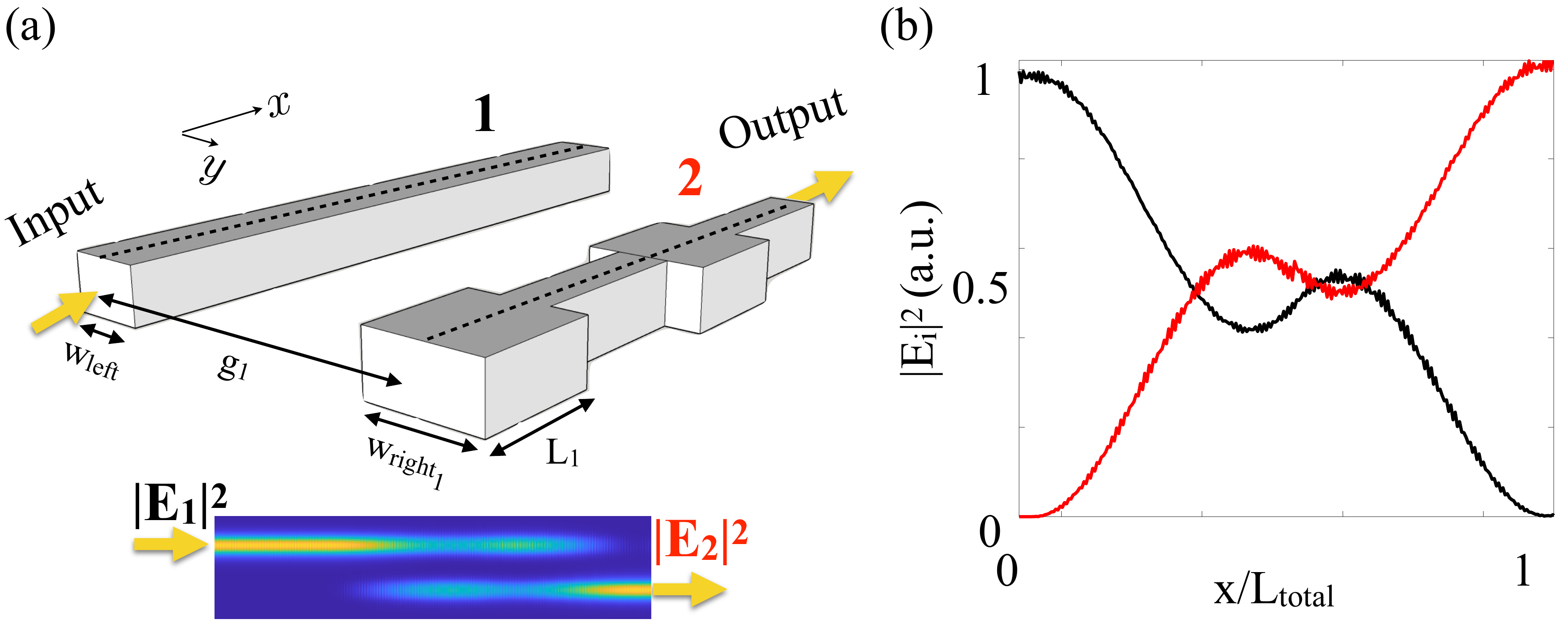}
\caption{\textbf{Complete light switching in a universal detuning-modulated composite waveguide coupler.} (a) (top) An out-of-scale schematic of the \emph{first-order} $N=4$-piece waveguide design. Light is initially injected to waveguide 1 and is transferred to waveguide 2. (bottom) Top-view of an eigenmode expansion solver (EME) calculation of the light intensity of the coupled waveguide system. (b) A cutline of the electromagnetic field intensity in the middle of waveguide 1 (black, initially populated) and waveguide 2 (red, initially empty) vs the normalized propagation length.
\label{Figure:Figure7}
}
\end{figure}

\section{Realization in integrated photonic circuits}
As detuning-modulated CPs enable implementation in systems with real-valued couplings, it is straightforward to realize the above universal sequences in quantum integrated photonic circuits \cite{wangIntPhot,TerryOptimistic}, which suffer from inaccuracies originating from fabrication defects and environmental characteristics (i.e. temperature). Universal DMCPs not only allow to overcome inevitable fabrication errors in such integrated devices, but relax the need for a precise initial state of light coupled into the system in order to achieve accurate gate operations. 

Fig. \ref{Figure:Figure7}(a) schematically shows a photonic circuit comprised of two coupled optical waveguides, situated at a distance of $g$ from their center lines. The amplitudes of the fundamental modes in the waveguides follow the coupled mode equations \cite{DirectionalWGs1}. Similar to Eq. \ref{Sch}, the coupling is $\Omega=ae^{-bg}$, where $a,b$ are material and geometry-dependent. The system is said to be on resonance if the two waveguides are identical in material and geometry, otherwise there exists a real-valued phase mismatch between the propagation constants $\beta_{i}$ which we define as the detuning $\Delta = (\beta_{1}-\beta_{2})/2$. The universal DMCPs can be applied to such a system by varying the relative widths of the waveguides to create discrete changes of $\Delta$ along the propagation axis. For further details of this realization, see Appendix B.

Fig. \ref{Figure:Figure7}(a) is a top view of the light intensity propagation of a $N=4$-piece coupled waveguide system that realizes the required changes in $\Delta$ to obtain a $\pi$ pulse (Table \ref{table:Table1}). This was simulated via an eigenmode expansion solver (EME) \cite{Lumerical}. The light intensity at the center line of each waveguide is plotted to stress the complete light switching in the coupled system in Fig. \ref{Figure:Figure7}(b). 

\section{Implementation in irreducible n-level systems with SU(2) symmetry}
The presented formalism for universal detuning modulated composite pulses is suitable for any irreducible n-level system with SU(2) symmetry. The notion of generalizing solutions from two-level systems to n-level systems with SU(2) symmetry may not be new \cite{PhysRevA.20.539, Hioe:87}, but it was recently adapted specifically to composite pulses \cite{doi:10.1063/1.5013672,shi2021robust}. One may consider non-degenerate levels, in which the diagonal elements of the Hamiltonian representing the system’s dynamics are the cumulative detunings of the excitation laser frequency from each Bohr frequency $\Delta_{n}$ and the off-diagonal elements link the different dipole transition moments between each two adjacent levels to the exciting electric field amplitude whose carrier frequency matches the Bohr frequency of this exact transition, namely $\Omega_{n}$ (i.e. the Rabi frequency for a transition between two adjacent levels). A variety of n-level solutions has been presented over the years, and in our manuscript, we refer to the Jacobi solution, given by $\Omega_{n} = \Omega_{0} \sqrt{n(N-n)}$ and $\Delta_{n} = n\Delta_{0} + D_{0}$. By using the irreducible matrix representation for SU(2) symmetry it was shown \cite{Hioe:87} that these off-diagonal elements are not necessarily equidistant. 

\begin{figure}
    \centering
    \includegraphics[scale=0.25]{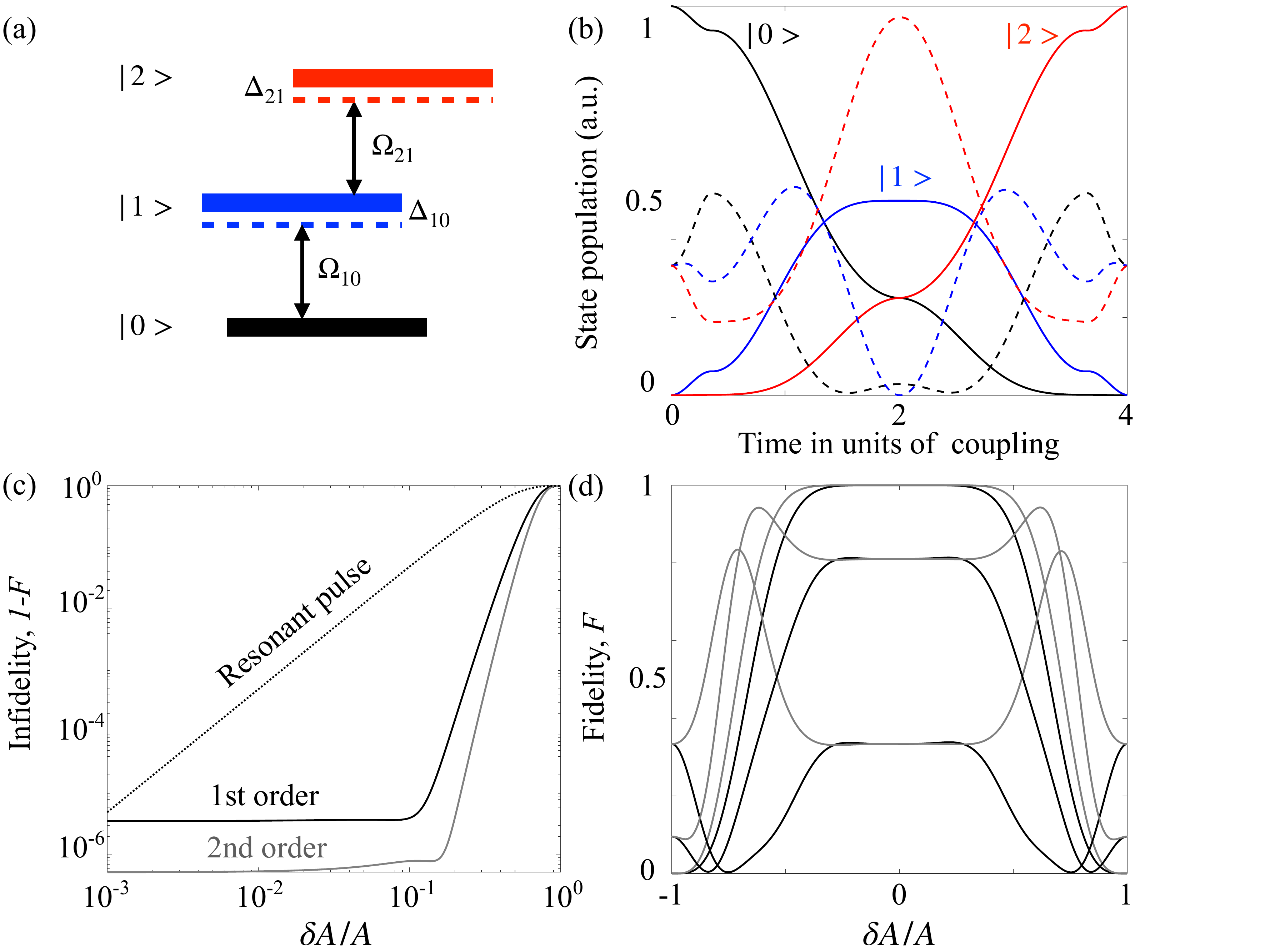}
    \caption{\textbf{Implementation of universal DMCPs in n-level systems with SU(2) symmetry.} (a) A $n=$ three-level irreducible system with Rabi frequencies $\Omega_{i,i+1}$ and detunings $\Delta_{i,i+1}$. (b) Palindromic state transfer from $|0>$ to $|2>$ in a three-level system  due to a first-order universal $N=4$ detuning-modulated composite $\pi$ pulse. In dashed lines, we demonstrate the state population for this pulse from an initial state that is an equal superposition of all states. (c) Infidelity (1-F) of the above first-order (black) and second-order (grey) $\pi$ pulse in logarithmic scale to errors in the target pulse area. Both outperform a single resonant pulse, shown for comparison, maintaining fidelity that is higher than the threshold for quantum information processing up to errors of $\delta A/A = 18\%$ for the first-order case and $\delta A/A = 27\%$ for the second-order case. (d) Fidelity (F) of the shortest first-(black lines) and second-order (grey lines) universal detuning-modulated composite $\pi$ pulses to errors in the target pulse area for different initial states: $|0\rangle$, $1/\sqrt{3}(|0 \rangle +|1\rangle +|2\rangle)$ and $0.9|0 \rangle +\sqrt{0.19/2}(|1\rangle +|2\rangle)$.}
    \label{fig:NLevelCP_1}
\end{figure}

Taking note of the three-level system in Fig. \ref{fig:NLevelCP_1}(a), of the two level pairs can be illuminated and excited with one of three fields: $\epsilon_{L1}(t) = A_{L1}e^{i\omega_{L1}t}$ with detuning $\Delta_{10} = \omega_{L1} - \omega_{10}$ or $\epsilon_{L2}(t) = A_{L2}e^{i\omega_{L2}t}$ with detuning $\Delta_{21} = \omega_{L2} - \omega_{21}$. In this case, the Rabi frequencies are $\Omega_{10} = 2A_{L1}(t)d_{10}/\hbar$ and $\Omega_{21} = 2A_{L2}(t)d_{21}/\hbar$. Assuming the above-mentioned Jacobi solution, and choosing $D_{0}=0$, one could set the excitation laser frequencies, such that the detunings are $\Delta_{10}=\Delta_{0}$ and $\Delta_{21}=2\Delta_{0}$. One can also chose the values of $A$ to comply with the Rabi frequencies in this solution, which will be equal in this case $\Omega_{10} = \Omega_{21}=\sqrt{2}\Omega_{0}$. 

In the following, we present an implementation of the above universal detuning-modulated composite pulses method to control a three-level system described by the SU(2) Hamiltonian:

\begin{equation}
    H = Re{\Omega_{g}}\hat{\sigma_{x}} + Im{\Omega_{g}}\hat{\sigma_{y}} + \Delta \hat{\sigma_{z}}
\end{equation}

where $\sigma_{i}$ are the $3\times3$ Pauli matrices. Assuming a $N=4$-piece universal detuning-modulated composite $\pi$ pulse, the system’s evolution is described by the composite matrix $U^{(N=4)} = U_{4}U_{3}U_{2}U_{1}$, where $U_{i}$, described by eq. \ref{U}, is the unitary propagator or the piece-wise Hamiltonian $H_{i} = Re{\Omega_{g,i}}\hat{\sigma_{x}} + Im{\Omega_{g,i}}\hat{\sigma_{y}} + \Delta_{i} \hat{\sigma_{z}}$.

For the sake of simplicity, we will set a constant coupling $\Omega_{i} = \Omega$, such that the complete composite sequence is governed by the detuning parameters derived in the above section. We reprise the shortest \emph{first-order} universal detuning modulated composite solution $\Delta = {[5.52,0.69,-0.69,-5.52]}\Omega$ and implement it on a three-level SU(2) symmetric system. This results in the state transfer shown in Fig. \ref{fig:NLevelCP_1}(b) that is robust to errors in the target pulse area from any initial state (Fig. \ref{fig:NLevelCP_1}(c-d)). The contour plot describing the robustness of this pulse sequence to errors in the target coupling and detuning parameters is similar to that of a two-level system. We note that as the unitary matrix elements grow in complexity for higher-level systems ($n \geq 4$), the shortest second-order sequence does not always outperform the first. That being said, all sequences maintain high fidelity against pulse area errors above the QIP threshold of $10^{-4}$.

Systems that portray SU(2) symmetry enable one to express the Hamiltonian of the n-level system in a compact and generalized form, and generally the levels are not constrained to be equal. Many physical systems portray these symmetries (e.g. topological insulators and $p_{x}+ip_{y}$ superconductors), thus the conditions of our derivations may be useful in the future for exciting such systems. 

\section{Conclusions} 
We introduce a new family of universal DMCPs to enable high-fidelity state transfer that is independent of the initial state of the system. These maintain robustness to pulse area errors within the quantum error threshold of $10^{-4}$ for n-level systems with SU(2) symmetry. We also show that our method offers a minimal pulse overhead and robustness to the system's lifetime, where the shortest sequence is composed of $N=4$ pulses. This technique is suitable for, but not limited to, implementation in integrated photonic circuits, which are considered a strong candidate for quantum computation hardware and QIP. As integrated photonic circuits are prone to fabrication errors, leading to a decrease in the produced signal fidelity, their application in QIP has been considered limited. Furthermore, precise quantum state preparation in integrated photonics requires an additional preliminary process of state preparation, which adds to the complexity of scaling-up device fabrication. To this end, universal detuning-modulated composite pulses enable the production of high fidelity light transfer without rigid requirements of the input signal. This versatile method will pave the way to achieve high precision in quantum gates via short and elegant solutions. It will further push the capabilities of reaching high fidelity in any qubit architecture and for large scale quantum circuits designs.

\bibliography{universalbib}

\begin{thebibliography}{36}%
\makeatletter
\providecommand \@ifxundefined [1]{%
 \@ifx{#1\undefined}
}%
\providecommand \@ifnum [1]{%
 \ifnum #1\expandafter \@firstoftwo
 \else \expandafter \@secondoftwo
 \fi
}%
\providecommand \@ifx [1]{%
 \ifx #1\expandafter \@firstoftwo
 \else \expandafter \@secondoftwo
 \fi
}%
\providecommand \natexlab [1]{#1}%
\providecommand \enquote  [1]{``#1''}%
\providecommand \bibnamefont  [1]{#1}%
\providecommand \bibfnamefont [1]{#1}%
\providecommand \citenamefont [1]{#1}%
\providecommand \href@noop [0]{\@secondoftwo}%
\providecommand \href [0]{\begingroup \@sanitize@url \@href}%
\providecommand \@href[1]{\@@startlink{#1}\@@href}%
\providecommand \@@href[1]{\endgroup#1\@@endlink}%
\providecommand \@sanitize@url [0]{\catcode `\\12\catcode `\$12\catcode
  `\&12\catcode `\#12\catcode `\^12\catcode `\_12\catcode `\%12\relax}%
\providecommand \@@startlink[1]{}%
\providecommand \@@endlink[0]{}%
\providecommand \url  [0]{\begingroup\@sanitize@url \@url }%
\providecommand \@url [1]{\endgroup\@href {#1}{\urlprefix }}%
\providecommand \urlprefix  [0]{URL }%
\providecommand \Eprint [0]{\href }%
\providecommand \doibase [0]{http://dx.doi.org/}%
\providecommand \selectlanguage [0]{\@gobble}%
\providecommand \bibinfo  [0]{\@secondoftwo}%
\providecommand \bibfield  [0]{\@secondoftwo}%
\providecommand \translation [1]{[#1]}%
\providecommand \BibitemOpen [0]{}%
\providecommand \bibitemStop [0]{}%
\providecommand \bibitemNoStop [0]{.\EOS\space}%
\providecommand \EOS [0]{\spacefactor3000\relax}%
\providecommand \BibitemShut  [1]{\csname bibitem#1\endcsname}%
\let\auto@bib@innerbib\@empty
%</preamble>
\bibitem [{\citenamefont {Nielsen}\ and\ \citenamefont
  {Chuang}(2000)}]{QuantumInfo1}%
  \BibitemOpen
  \bibfield  {author} {\bibinfo {author} {\bibfnamefont {M.}~\bibnamefont
  {Nielsen}}\ and\ \bibinfo {author} {\bibfnamefont {I.}~\bibnamefont
  {Chuang}},\ }\href {https://doi.org/10.1017/CBO9780511976667} {\emph
  {\bibinfo {title} {Quantum Computation and Quantum Information}}}\ (\bibinfo
  {publisher} {Cambridge University Press},\ \bibinfo {address} {Cambridge},\
  \bibinfo {year} {2000})\BibitemShut {NoStop}%
\bibitem [{\citenamefont {Zener}(1932)}]{Zener}%
  \BibitemOpen
  \bibfield  {author} {\bibinfo {author} {\bibfnamefont {C.}~\bibnamefont
  {Zener}},\ }\href@noop {} {\bibfield  {journal} {\bibinfo  {journal} {Proc.
  R. Soc. London Ser. A}\ }\textbf {\bibinfo {volume} {137}},\ \bibinfo {pages}
  {696} (\bibinfo {year} {1932})}\BibitemShut {NoStop}%
\bibitem [{\citenamefont {Rosen}\ and\ \citenamefont
  {Zener}(1932)}]{RosenZener}%
  \BibitemOpen
  \bibfield  {author} {\bibinfo {author} {\bibfnamefont {N.}~\bibnamefont
  {Rosen}}\ and\ \bibinfo {author} {\bibfnamefont {C.}~\bibnamefont {Zener}},\
  }\href@noop {} {\bibfield  {journal} {\bibinfo  {journal} {Phys. Rev.}\
  }\textbf {\bibinfo {volume} {40}},\ \bibinfo {pages} {502} (\bibinfo {year}
  {1932})}\BibitemShut {NoStop}%
\bibitem [{\citenamefont {Allen}\ and\ \citenamefont
  {Eberly}(1975)}]{AllenEberly}%
  \BibitemOpen
  \bibfield  {author} {\bibinfo {author} {\bibfnamefont {L.}~\bibnamefont
  {Allen}}\ and\ \bibinfo {author} {\bibfnamefont {J.}~\bibnamefont {Eberly}},\
  }\href@noop {} {\emph {\bibinfo {title} {Optical Resonance and Two Level
  Systems}}}\ (\bibinfo  {publisher} {Dover},\ \bibinfo {address} {New York},\
  \bibinfo {year} {1975})\BibitemShut {NoStop}%
\bibitem [{\citenamefont {Gaubatz}\ \emph {et~al.}(1990)\citenamefont
  {Gaubatz}, \citenamefont {Rudecki}, \citenamefont {Schiemann},\ and\
  \citenamefont {Bergmann}}]{STIRAP}%
  \BibitemOpen
  \bibfield  {author} {\bibinfo {author} {\bibfnamefont {U.}~\bibnamefont
  {Gaubatz}}, \bibinfo {author} {\bibfnamefont {P.}~\bibnamefont {Rudecki}},
  \bibinfo {author} {\bibfnamefont {S.}~\bibnamefont {Schiemann}}, \ and\
  \bibinfo {author} {\bibfnamefont {K.}~\bibnamefont {Bergmann}},\ }\href
  {\doibase 10.1063/1.458514} {\bibfield  {journal} {\bibinfo  {journal} {The
  Journal of Chemical Physics}\ }\textbf {\bibinfo {volume} {92}},\ \bibinfo
  {pages} {5363} (\bibinfo {year} {1990})},\ \Eprint
  {http://arxiv.org/abs/https://doi.org/10.1063/1.458514}
  {https://doi.org/10.1063/1.458514} \BibitemShut {NoStop}%
\bibitem [{\citenamefont {Hahn}(1950)}]{CP1}%
  \BibitemOpen
  \bibfield  {author} {\bibinfo {author} {\bibfnamefont {E.~L.}\ \bibnamefont
  {Hahn}},\ }\href {\doibase 10.1103/PhysRev.80.580} {\bibfield  {journal}
  {\bibinfo  {journal} {Phys. Rev.}\ }\textbf {\bibinfo {volume} {80}},\
  \bibinfo {pages} {580} (\bibinfo {year} {1950})}\BibitemShut {NoStop}%
\bibitem [{\citenamefont {Carr}\ and\ \citenamefont
  {Purcell}(1954)}]{CarrSpinEcho}%
  \BibitemOpen
  \bibfield  {author} {\bibinfo {author} {\bibfnamefont {H.~Y.}\ \bibnamefont
  {Carr}}\ and\ \bibinfo {author} {\bibfnamefont {E.~M.}\ \bibnamefont
  {Purcell}},\ }\href {\doibase 10.1103/PhysRev.94.630} {\bibfield  {journal}
  {\bibinfo  {journal} {Phys. Rev.}\ }\textbf {\bibinfo {volume} {94}},\
  \bibinfo {pages} {630} (\bibinfo {year} {1954})}\BibitemShut {NoStop}%
\bibitem [{\citenamefont {Cho}\ \emph {et~al.}(1986)\citenamefont {Cho},
  \citenamefont {Tycko}, \citenamefont {Pines},\ and\ \citenamefont
  {Guckenheimer}}]{CP2}%
  \BibitemOpen
  \bibfield  {author} {\bibinfo {author} {\bibfnamefont {H.~M.}\ \bibnamefont
  {Cho}}, \bibinfo {author} {\bibfnamefont {R.}~\bibnamefont {Tycko}}, \bibinfo
  {author} {\bibfnamefont {A.}~\bibnamefont {Pines}}, \ and\ \bibinfo {author}
  {\bibfnamefont {J.}~\bibnamefont {Guckenheimer}},\ }\href {\doibase
  10.1103/PhysRevLett.56.1905} {\bibfield  {journal} {\bibinfo  {journal}
  {Phys. Rev. Lett.}\ }\textbf {\bibinfo {volume} {56}},\ \bibinfo {pages}
  {1905} (\bibinfo {year} {1986})}\BibitemShut {NoStop}%
\bibitem [{\citenamefont {Cummins}\ \emph {et~al.}(2003)\citenamefont
  {Cummins}, \citenamefont {Llewellyn},\ and\ \citenamefont {Jones}}]{CP3}%
  \BibitemOpen
  \bibfield  {author} {\bibinfo {author} {\bibfnamefont {H.~K.}\ \bibnamefont
  {Cummins}}, \bibinfo {author} {\bibfnamefont {G.}~\bibnamefont {Llewellyn}},
  \ and\ \bibinfo {author} {\bibfnamefont {J.~A.}\ \bibnamefont {Jones}},\
  }\href {\doibase 10.1103/PhysRevA.67.042308} {\bibfield  {journal} {\bibinfo
  {journal} {Phys. Rev. A}\ }\textbf {\bibinfo {volume} {67}},\ \bibinfo
  {pages} {042308} (\bibinfo {year} {2003})}\BibitemShut {NoStop}%
\bibitem [{\citenamefont {Levitt}\ and\ \citenamefont {Freeman}(1979)}]{CP4}%
  \BibitemOpen
  \bibfield  {author} {\bibinfo {author} {\bibfnamefont {M.~H.}\ \bibnamefont
  {Levitt}}\ and\ \bibinfo {author} {\bibfnamefont {R.}~\bibnamefont
  {Freeman}},\ }\href {https://doi.org/10.1016/0022-2364(79)90265-8} {\bibfield
   {journal} {\bibinfo  {journal} {J. Magn. Reson.}\ }\textbf {\bibinfo
  {volume} {33}},\ \bibinfo {pages} {473} (\bibinfo {year} {1979})}\BibitemShut
  {NoStop}%
\bibitem [{\citenamefont {Freeman}\ \emph {et~al.}(1980)\citenamefont
  {Freeman}, \citenamefont {Kempsell},\ and\ \citenamefont {Levitt}}]{CP5}%
  \BibitemOpen
  \bibfield  {author} {\bibinfo {author} {\bibfnamefont {R.}~\bibnamefont
  {Freeman}}, \bibinfo {author} {\bibfnamefont {S.}~\bibnamefont {Kempsell}}, \
  and\ \bibinfo {author} {\bibfnamefont {M.}~\bibnamefont {Levitt}},\ }\href
  {https://doi.org/10.1016/0022-2364(80)90327-3} {\bibfield  {journal}
  {\bibinfo  {journal} {J. Magn. Reson.}\ }\textbf {\bibinfo {volume} {38}},\
  \bibinfo {pages} {453} (\bibinfo {year} {1980})}\BibitemShut {NoStop}%
\bibitem [{\citenamefont {Levitt}(1986)}]{CP6}%
  \BibitemOpen
  \bibfield  {author} {\bibinfo {author} {\bibfnamefont {M.}~\bibnamefont
  {Levitt}},\ }\href {https://doi.org/10.1016/0079-6565(86)80005-X} {\bibfield
  {journal} {\bibinfo  {journal} {Prog. Nucl. Magn. Reson. Spectrosc.}\
  }\textbf {\bibinfo {volume} {18}},\ \bibinfo {pages} {61} (\bibinfo {year}
  {1986})}\BibitemShut {NoStop}%
\bibitem [{\citenamefont {Tycko}(1983)}]{CP7}%
  \BibitemOpen
  \bibfield  {author} {\bibinfo {author} {\bibfnamefont {R.}~\bibnamefont
  {Tycko}},\ }\href {\doibase 10.1103/PhysRevLett.51.775} {\bibfield  {journal}
  {\bibinfo  {journal} {Phys. Rev. Lett.}\ }\textbf {\bibinfo {volume} {51}},\
  \bibinfo {pages} {775} (\bibinfo {year} {1983})}\BibitemShut {NoStop}%
\bibitem [{\citenamefont {Torosov}\ and\ \citenamefont
  {Vitanov}(2011)}]{VitanovSmooth}%
  \BibitemOpen
  \bibfield  {author} {\bibinfo {author} {\bibfnamefont {B.~T.}\ \bibnamefont
  {Torosov}}\ and\ \bibinfo {author} {\bibfnamefont {N.~V.}\ \bibnamefont
  {Vitanov}},\ }\href {\doibase 10.1103/PhysRevA.83.053420} {\bibfield
  {journal} {\bibinfo  {journal} {Phys. Rev. A}\ }\textbf {\bibinfo {volume}
  {83}},\ \bibinfo {pages} {053420} (\bibinfo {year} {2011})}\BibitemShut
  {NoStop}%
\bibitem [{\citenamefont {Genov}\ \emph {et~al.}(2014)\citenamefont {Genov},
  \citenamefont {Schraft}, \citenamefont {Halfmann},\ and\ \citenamefont
  {Vitanov}}]{GenovPRL}%
  \BibitemOpen
  \bibfield  {author} {\bibinfo {author} {\bibfnamefont {G.~T.}\ \bibnamefont
  {Genov}}, \bibinfo {author} {\bibfnamefont {D.}~\bibnamefont {Schraft}},
  \bibinfo {author} {\bibfnamefont {T.}~\bibnamefont {Halfmann}}, \ and\
  \bibinfo {author} {\bibfnamefont {N.~V.}\ \bibnamefont {Vitanov}},\ }\href
  {\doibase 10.1103/PhysRevLett.113.043001} {\bibfield  {journal} {\bibinfo
  {journal} {Phys. Rev. Lett.}\ }\textbf {\bibinfo {volume} {113}},\ \bibinfo
  {pages} {043001} (\bibinfo {year} {2014})}\BibitemShut {NoStop}%
\bibitem [{\citenamefont {Keeler}(2005)}]{KeelerNMR}%
  \BibitemOpen
  \bibfield  {author} {\bibinfo {author} {\bibfnamefont {J.}~\bibnamefont
  {Keeler}},\ }\href {https://doi.org/10.17863/CAM.968} {\emph {\bibinfo
  {title} {Understanding NMR Spectroscopy}}}\ (\bibinfo  {publisher} {John
  Wiley and Sons},\ \bibinfo {address} {New York},\ \bibinfo {year}
  {2005})\BibitemShut {NoStop}%
\bibitem [{\citenamefont {Jones}(2009)}]{jones2009composite}%
  \BibitemOpen
  \bibfield  {author} {\bibinfo {author} {\bibfnamefont {J.~A.}\ \bibnamefont
  {Jones}},\ }\href@noop {} {\bibfield  {journal} {\bibinfo  {journal} {arXiv
  preprint arXiv:0906.4719}\ } (\bibinfo {year} {2009})}\BibitemShut {NoStop}%
\bibitem [{\citenamefont {Husain}\ \emph {et~al.}(2013)\citenamefont {Husain},
  \citenamefont {Kawamura},\ and\ \citenamefont {Jones}}]{husain2013further}%
  \BibitemOpen
  \bibfield  {author} {\bibinfo {author} {\bibfnamefont {S.}~\bibnamefont
  {Husain}}, \bibinfo {author} {\bibfnamefont {M.}~\bibnamefont {Kawamura}}, \
  and\ \bibinfo {author} {\bibfnamefont {J.~A.}\ \bibnamefont {Jones}},\
  }\href@noop {} {\bibfield  {journal} {\bibinfo  {journal} {Journal of
  Magnetic Resonance}\ }\textbf {\bibinfo {volume} {230}},\ \bibinfo {pages}
  {145} (\bibinfo {year} {2013})}\BibitemShut {NoStop}%
\bibitem [{\citenamefont {Alexander}\ \emph {et~al.}(2020)\citenamefont
  {Alexander}, \citenamefont {Kanazawa}, \citenamefont {Egger}, \citenamefont
  {Capelluto}, \citenamefont {Wood}, \citenamefont {Javadi-Abhari},\ and\
  \citenamefont {McKay}}]{alexander2020qiskit}%
  \BibitemOpen
  \bibfield  {author} {\bibinfo {author} {\bibfnamefont {T.}~\bibnamefont
  {Alexander}}, \bibinfo {author} {\bibfnamefont {N.}~\bibnamefont {Kanazawa}},
  \bibinfo {author} {\bibfnamefont {D.~J.}\ \bibnamefont {Egger}}, \bibinfo
  {author} {\bibfnamefont {L.}~\bibnamefont {Capelluto}}, \bibinfo {author}
  {\bibfnamefont {C.~J.}\ \bibnamefont {Wood}}, \bibinfo {author}
  {\bibfnamefont {A.}~\bibnamefont {Javadi-Abhari}}, \ and\ \bibinfo {author}
  {\bibfnamefont {D.~C.}\ \bibnamefont {McKay}},\ }\href@noop {} {\bibfield
  {journal} {\bibinfo  {journal} {Quantum Science and Technology}\ }\textbf
  {\bibinfo {volume} {5}},\ \bibinfo {pages} {044006} (\bibinfo {year}
  {2020})}\BibitemShut {NoStop}%
\bibitem [{\citenamefont {Kyoseva}\ \emph {et~al.}(2019)\citenamefont
  {Kyoseva}, \citenamefont {Greener},\ and\ \citenamefont
  {Suchowski}}]{PhysRevA.100.032333}%
  \BibitemOpen
  \bibfield  {author} {\bibinfo {author} {\bibfnamefont {E.}~\bibnamefont
  {Kyoseva}}, \bibinfo {author} {\bibfnamefont {H.}~\bibnamefont {Greener}}, \
  and\ \bibinfo {author} {\bibfnamefont {H.}~\bibnamefont {Suchowski}},\ }\href
  {\doibase 10.1103/PhysRevA.100.032333} {\bibfield  {journal} {\bibinfo
  {journal} {Phys. Rev. A}\ }\textbf {\bibinfo {volume} {100}},\ \bibinfo
  {pages} {032333} (\bibinfo {year} {2019})}\BibitemShut {NoStop}%
\bibitem [{\citenamefont {Kyoseva}\ and\ \citenamefont
  {Vitanov}(2013)}]{Elica:passband}%
  \BibitemOpen
  \bibfield  {author} {\bibinfo {author} {\bibfnamefont {E.}~\bibnamefont
  {Kyoseva}}\ and\ \bibinfo {author} {\bibfnamefont {N.~V.}\ \bibnamefont
  {Vitanov}},\ }\href {\doibase 10.1103/PhysRevA.88.063410} {\bibfield
  {journal} {\bibinfo  {journal} {Phys. Rev. A}\ }\textbf {\bibinfo {volume}
  {88}},\ \bibinfo {pages} {063410} (\bibinfo {year} {2013})}\BibitemShut
  {NoStop}%
\bibitem [{\citenamefont {Luy}\ \emph {et~al.}(2005)\citenamefont {Luy},
  \citenamefont {Kobzar}, \citenamefont {Skinner}, \citenamefont {Khaneja},\
  and\ \citenamefont {Glaser}}]{LUY2005179}%
  \BibitemOpen
  \bibfield  {author} {\bibinfo {author} {\bibfnamefont {B.}~\bibnamefont
  {Luy}}, \bibinfo {author} {\bibfnamefont {K.}~\bibnamefont {Kobzar}},
  \bibinfo {author} {\bibfnamefont {T.~E.}\ \bibnamefont {Skinner}}, \bibinfo
  {author} {\bibfnamefont {N.}~\bibnamefont {Khaneja}}, \ and\ \bibinfo
  {author} {\bibfnamefont {S.~J.}\ \bibnamefont {Glaser}},\ }\href {\doibase
  https://doi.org/10.1016/j.jmr.2005.06.002} {\bibfield  {journal} {\bibinfo
  {journal} {Journal of Magnetic Resonance}\ }\textbf {\bibinfo {volume}
  {176}},\ \bibinfo {pages} {179 } (\bibinfo {year} {2005})}\BibitemShut
  {NoStop}%
\bibitem [{\citenamefont {Frimmer}\ and\ \citenamefont
  {Novotny}(2014)}]{relaxation1}%
  \BibitemOpen
  \bibfield  {author} {\bibinfo {author} {\bibfnamefont {M.}~\bibnamefont
  {Frimmer}}\ and\ \bibinfo {author} {\bibfnamefont {L.}~\bibnamefont
  {Novotny}},\ }\href {\doibase 10.1119/1.4878621} {\bibfield  {journal}
  {\bibinfo  {journal} {American Journal of Physics}\ }\textbf {\bibinfo
  {volume} {82}},\ \bibinfo {pages} {947} (\bibinfo {year} {2014})},\ \Eprint
  {http://arxiv.org/abs/https://doi.org/10.1119/1.4878621}
  {https://doi.org/10.1119/1.4878621} \BibitemShut {NoStop}%
\bibitem [{\citenamefont {Falci}\ \emph {et~al.}(2005)\citenamefont {Falci},
  \citenamefont {D'Arrigo}, \citenamefont {Mastellone},\ and\ \citenamefont
  {Paladino}}]{relaxation2}%
  \BibitemOpen
  \bibfield  {author} {\bibinfo {author} {\bibfnamefont {G.}~\bibnamefont
  {Falci}}, \bibinfo {author} {\bibfnamefont {A.}~\bibnamefont {D'Arrigo}},
  \bibinfo {author} {\bibfnamefont {A.}~\bibnamefont {Mastellone}}, \ and\
  \bibinfo {author} {\bibfnamefont {E.}~\bibnamefont {Paladino}},\ }\href
  {\doibase 10.1103/PhysRevLett.94.167002} {\bibfield  {journal} {\bibinfo
  {journal} {Phys. Rev. Lett.}\ }\textbf {\bibinfo {volume} {94}},\ \bibinfo
  {pages} {167002} (\bibinfo {year} {2005})}\BibitemShut {NoStop}%
\bibitem [{\citenamefont {Collin}\ \emph {et~al.}(2004)\citenamefont {Collin},
  \citenamefont {Ithier}, \citenamefont {Aassime}, \citenamefont {Joyez},
  \citenamefont {Vion},\ and\ \citenamefont {Esteve}}]{PhysRevLett.93.157005}%
  \BibitemOpen
  \bibfield  {author} {\bibinfo {author} {\bibfnamefont {E.}~\bibnamefont
  {Collin}}, \bibinfo {author} {\bibfnamefont {G.}~\bibnamefont {Ithier}},
  \bibinfo {author} {\bibfnamefont {A.}~\bibnamefont {Aassime}}, \bibinfo
  {author} {\bibfnamefont {P.}~\bibnamefont {Joyez}}, \bibinfo {author}
  {\bibfnamefont {D.}~\bibnamefont {Vion}}, \ and\ \bibinfo {author}
  {\bibfnamefont {D.}~\bibnamefont {Esteve}},\ }\href {\doibase
  10.1103/PhysRevLett.93.157005} {\bibfield  {journal} {\bibinfo  {journal}
  {Phys. Rev. Lett.}\ }\textbf {\bibinfo {volume} {93}},\ \bibinfo {pages}
  {157005} (\bibinfo {year} {2004})}\BibitemShut {NoStop}%
\bibitem [{\citenamefont {Faust}\ \emph {et~al.}(2013)\citenamefont {Faust},
  \citenamefont {Rieger}, \citenamefont {Seitner}, \citenamefont {Kotthaus},\
  and\ \citenamefont {Weig}}]{Decoherence1}%
  \BibitemOpen
  \bibfield  {author} {\bibinfo {author} {\bibfnamefont {T.}~\bibnamefont
  {Faust}}, \bibinfo {author} {\bibfnamefont {J.}~\bibnamefont {Rieger}},
  \bibinfo {author} {\bibfnamefont {M.~J.}\ \bibnamefont {Seitner}}, \bibinfo
  {author} {\bibfnamefont {J.~P.}\ \bibnamefont {Kotthaus}}, \ and\ \bibinfo
  {author} {\bibfnamefont {E.~M.}\ \bibnamefont {Weig}},\ }\href
  {https://doi.org/10.1038/nphys2666} {\bibfield  {journal} {\bibinfo
  {journal} {Nature Physics}\ }\textbf {\bibinfo {volume} {9}},\ \bibinfo
  {pages} {485 EP } (\bibinfo {year} {2013})}\BibitemShut {NoStop}%
\bibitem [{\citenamefont {Hanson}\ \emph {et~al.}(2006)\citenamefont {Hanson},
  \citenamefont {Gywat},\ and\ \citenamefont {Awschalom}}]{PhysRevB.74.161203}%
  \BibitemOpen
  \bibfield  {author} {\bibinfo {author} {\bibfnamefont {R.}~\bibnamefont
  {Hanson}}, \bibinfo {author} {\bibfnamefont {O.}~\bibnamefont {Gywat}}, \
  and\ \bibinfo {author} {\bibfnamefont {D.~D.}\ \bibnamefont {Awschalom}},\
  }\href {\doibase 10.1103/PhysRevB.74.161203} {\bibfield  {journal} {\bibinfo
  {journal} {Phys. Rev. B}\ }\textbf {\bibinfo {volume} {74}},\ \bibinfo
  {pages} {161203} (\bibinfo {year} {2006})}\BibitemShut {NoStop}%
\bibitem [{\citenamefont {Takahashi}\ \emph {et~al.}(2012)\citenamefont
  {Takahashi}, \citenamefont {Brunel}, \citenamefont {Edwards}, \citenamefont
  {van Tol}, \citenamefont {Ramian}, \citenamefont {Han},\ and\ \citenamefont
  {Sherwin}}]{Decoherence2}%
  \BibitemOpen
  \bibfield  {author} {\bibinfo {author} {\bibfnamefont {S.}~\bibnamefont
  {Takahashi}}, \bibinfo {author} {\bibfnamefont {L.~C.}\ \bibnamefont
  {Brunel}}, \bibinfo {author} {\bibfnamefont {D.~T.}\ \bibnamefont {Edwards}},
  \bibinfo {author} {\bibfnamefont {J.}~\bibnamefont {van Tol}}, \bibinfo
  {author} {\bibfnamefont {G.}~\bibnamefont {Ramian}}, \bibinfo {author}
  {\bibfnamefont {S.}~\bibnamefont {Han}}, \ and\ \bibinfo {author}
  {\bibfnamefont {M.~S.}\ \bibnamefont {Sherwin}},\ }\href {\doibase
  10.1038/nature11437} {\bibfield  {journal} {\bibinfo  {journal} {Nature}\
  }\textbf {\bibinfo {volume} {489}},\ \bibinfo {pages} {409} (\bibinfo {year}
  {2012})}\BibitemShut {NoStop}%
\bibitem [{\citenamefont {Wang}\ \emph {et~al.}(2020)\citenamefont {Wang},
  \citenamefont {Sciarrino}, \citenamefont {Laing},\ and\ \citenamefont
  {Thompson}}]{wangIntPhot}%
  \BibitemOpen
  \bibfield  {author} {\bibinfo {author} {\bibfnamefont {J.}~\bibnamefont
  {Wang}}, \bibinfo {author} {\bibfnamefont {F.}~\bibnamefont {Sciarrino}},
  \bibinfo {author} {\bibfnamefont {A.}~\bibnamefont {Laing}}, \ and\ \bibinfo
  {author} {\bibfnamefont {M.~G.}\ \bibnamefont {Thompson}},\ }\href {\doibase
  10.1038/s41566-019-0532-1} {\bibfield  {journal} {\bibinfo  {journal} {Nature
  Photonics}\ }\textbf {\bibinfo {volume} {14}},\ \bibinfo {pages} {273}
  (\bibinfo {year} {2020})}\BibitemShut {NoStop}%
\bibitem [{\citenamefont {Rudolph}(2017)}]{TerryOptimistic}%
  \BibitemOpen
  \bibfield  {author} {\bibinfo {author} {\bibfnamefont {T.}~\bibnamefont
  {Rudolph}},\ }\href {\doibase 10.1063/1.4976737} {\bibfield  {journal}
  {\bibinfo  {journal} {APL Photonics}\ }\textbf {\bibinfo {volume} {2}},\
  \bibinfo {pages} {030901} (\bibinfo {year} {2017})},\ \Eprint
  {http://arxiv.org/abs/https://doi.org/10.1063/1.4976737}
  {https://doi.org/10.1063/1.4976737} \BibitemShut {NoStop}%
\bibitem [{\citenamefont {Yariv}(1973)}]{DirectionalWGs1}%
  \BibitemOpen
  \bibfield  {author} {\bibinfo {author} {\bibfnamefont {A.}~\bibnamefont
  {Yariv}},\ }\href {https://doi.org/10.1002/piuz.19760070610} {\bibfield
  {journal} {\bibinfo  {journal} {IEEE, J. Quantum Electronics}\ }\textbf
  {\bibinfo {volume} {9}},\ \bibinfo {pages} {9} (\bibinfo {year}
  {1973})}\BibitemShut {NoStop}%
\bibitem [{Lum()}]{Lumerical}%
  \BibitemOpen
  \href@noop {} {\enquote {\bibinfo {title} {Lumerical inc.
  https://www.lumerical.com/products/},}\ }\BibitemShut {NoStop}%
\bibitem [{\citenamefont {Cook}\ and\ \citenamefont
  {Shore}(1979)}]{PhysRevA.20.539}%
  \BibitemOpen
  \bibfield  {author} {\bibinfo {author} {\bibfnamefont {R.~J.}\ \bibnamefont
  {Cook}}\ and\ \bibinfo {author} {\bibfnamefont {B.~W.}\ \bibnamefont
  {Shore}},\ }\href {\doibase 10.1103/PhysRevA.20.539} {\bibfield  {journal}
  {\bibinfo  {journal} {Phys. Rev. A}\ }\textbf {\bibinfo {volume} {20}},\
  \bibinfo {pages} {539} (\bibinfo {year} {1979})}\BibitemShut {NoStop}%
\bibitem [{\citenamefont {Hioe}(1987)}]{Hioe:87}%
  \BibitemOpen
  \bibfield  {author} {\bibinfo {author} {\bibfnamefont {F.~T.}\ \bibnamefont
  {Hioe}},\ }\href {\doibase 10.1364/JOSAB.4.001327} {\bibfield  {journal}
  {\bibinfo  {journal} {J. Opt. Soc. Am. B}\ }\textbf {\bibinfo {volume} {4}},\
  \bibinfo {pages} {1327} (\bibinfo {year} {1987})}\BibitemShut {NoStop}%
\bibitem [{\citenamefont {Greener}\ and\ \citenamefont
  {Suchowski}(2018)}]{doi:10.1063/1.5013672}%
  \BibitemOpen
  \bibfield  {author} {\bibinfo {author} {\bibfnamefont {H.}~\bibnamefont
  {Greener}}\ and\ \bibinfo {author} {\bibfnamefont {H.}~\bibnamefont
  {Suchowski}},\ }\href {\doibase 10.1063/1.5013672} {\bibfield  {journal}
  {\bibinfo  {journal} {The Journal of Chemical Physics}\ }\textbf {\bibinfo
  {volume} {148}},\ \bibinfo {pages} {074101} (\bibinfo {year}
  {2018})}\BibitemShut {NoStop}%
\bibitem [{\citenamefont {Shi}\ \emph {et~al.}(2021)\citenamefont {Shi},
  \citenamefont {Wu}, \citenamefont {Shen}, \citenamefont {Song}, \citenamefont
  {Xia}, \citenamefont {Yi},\ and\ \citenamefont {Zheng}}]{shi2021robust}%
  \BibitemOpen
  \bibfield  {author} {\bibinfo {author} {\bibfnamefont {Z.-C.}\ \bibnamefont
  {Shi}}, \bibinfo {author} {\bibfnamefont {H.-N.}\ \bibnamefont {Wu}},
  \bibinfo {author} {\bibfnamefont {L.-T.}\ \bibnamefont {Shen}}, \bibinfo
  {author} {\bibfnamefont {J.}~\bibnamefont {Song}}, \bibinfo {author}
  {\bibfnamefont {Y.}~\bibnamefont {Xia}}, \bibinfo {author} {\bibfnamefont
  {X.}~\bibnamefont {Yi}}, \ and\ \bibinfo {author} {\bibfnamefont {S.-B.}\
  \bibnamefont {Zheng}},\ }\href@noop {} {\bibfield  {journal} {\bibinfo
  {journal} {Physical Review A}\ }\textbf {\bibinfo {volume} {103}},\ \bibinfo
  {pages} {052612} (\bibinfo {year} {2021})}\BibitemShut {NoStop}%
\end{thebibliography}%



\subsection*{Acknowledgements}
H. S. is supported by ERC-StG MIRAGE 20-15 project.

\appendix
\section{Fidelity of pulses against coupling and detuning errors}
Contrary to a standard single resonant pulse, which has only one point (zero dimensional) in parameter space that allows for accurate state transfer, in the composite designs, there is a one-dimensional parameter space that allows for the robustness of the composite pulses as a function of errors in the target parameters. As stated in the main text (see Fig. 3-4), these parameters are the individual target pulse areas $A_{i}$, and thus the target coupling $\Omega_{i}$ and target detuning $\Delta_{i}$ values. In this parameter space, there is a contour that allows for robustness to both target coupling and detuning errors that is above $10^{-4}$ fidelity. In the following, we provide the full-scaled contour plots for universal DMCP $\pi$ and $\pi/2$ pulses. Namely, given up to $100\%$ error in the target values of coupling and/or detuning, we measured the fidelity of the target state. Note that for a $\pi/2$ pulse, the target state is $\frac{1}{2}$.

\begin{figure}[h!]
\centering
\includegraphics[scale=0.30]{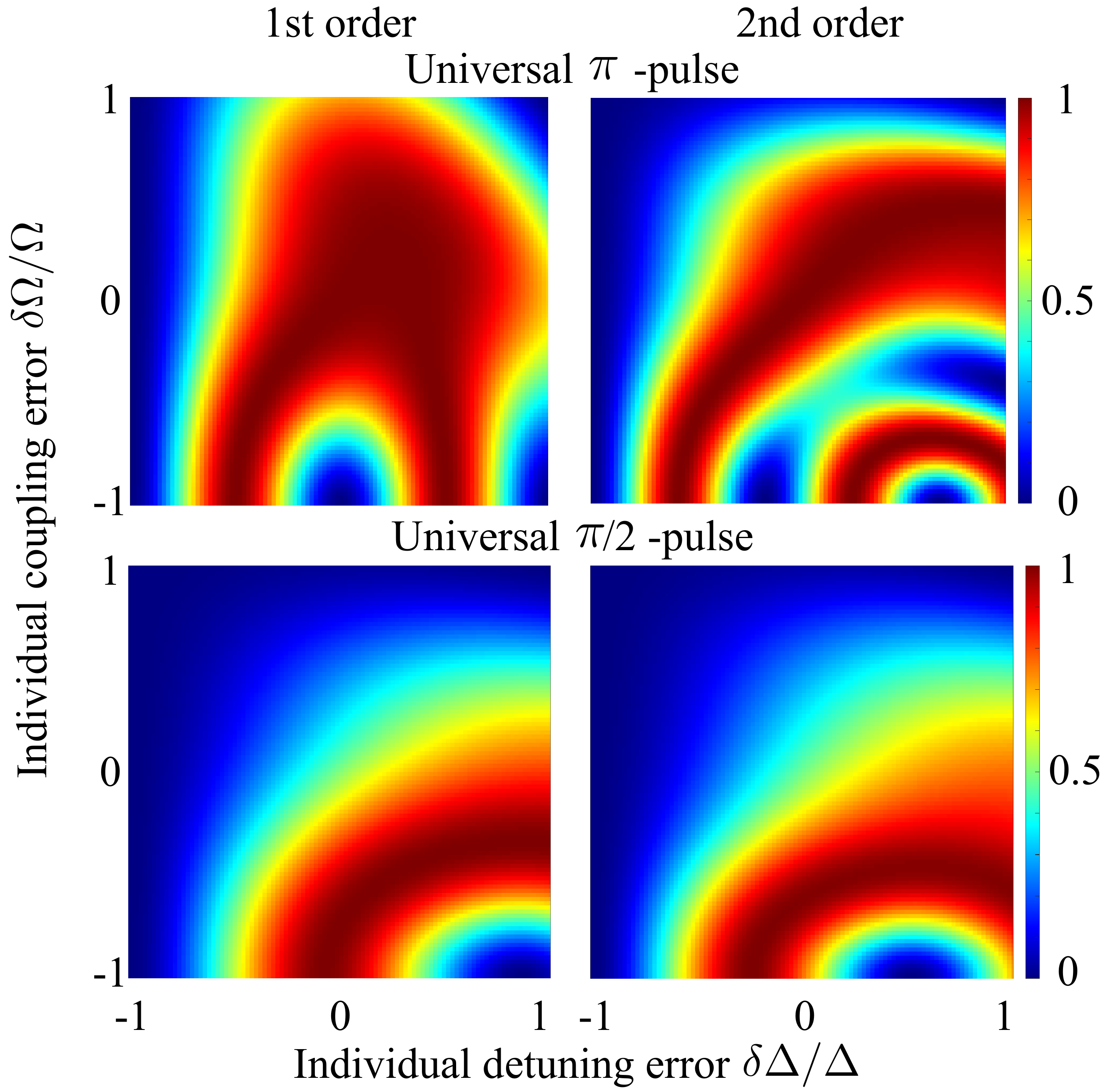}
\caption{\textbf{Contour plots of the fidelity of universal detuning-modulated pulses.} 1st-(left) and 2nd-order (right) universal $\pi$ (top) and $\pi/2$ (bottom) detuning-modulated pulses as a function of individual detuning $\delta \Delta/\Delta$ and coupling $\delta \Omega/\Omega$ errors. The contour plots show large areas in which the pulses are robust to errors in target values of these parameters (see text for details).
\label{Figure:SFig4}}
\end{figure}

\begin{figure*}
\centering
\includegraphics[scale=0.3]{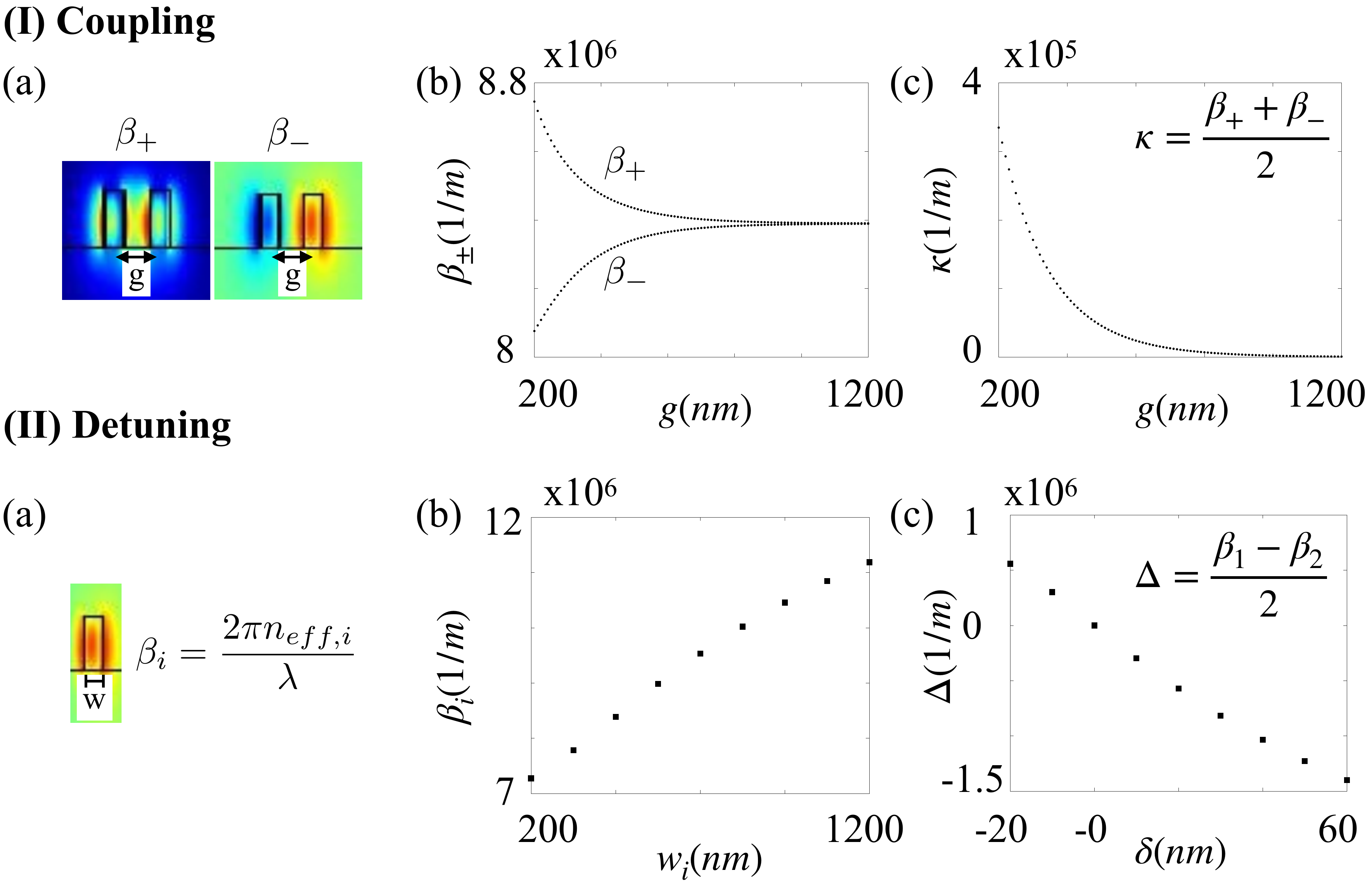}
\caption{\textbf{Coupling and detuning values of coupled waveguides.} (I) (a) A finite difference eigenmode solver (FDE) is used to calculate the coupling parameter $\kappa$ of two identical $Si$ on $SiO_{2}$ waveguides of width $220nm$ and height $340nm$ set apart at a distance $g$ between their centerlines. (b) The symmetric $\beta_{+}$ and antisymmetric $\beta_{-}$ modes of the coupled waveguides as a function of $g$ for an input wavelength of $\lambda = 1310nm$. (c) The resulting coupling parameter as a function of $g$. (II) (a) FDE calculation of the propagation constant of a single waveguide of width $w$. (b) The propagation constant of a single waveguide $\beta_{i}$ as a function of its width $w_{i}$. (c) The resulting detuning of a coupled waveguide system as a function of the difference in their widths $\delta$, where the width of waveguide 1 is $w$ and the width of waveguide 2 is $w \pm \delta$. 
\label{Figure:SFig2}}
\end{figure*}

\section{Schematic of realization in quantum integrated photonics}
In the following, we provide details on the scheme that translates the universal detuning-modulated CP theory to a realization in quantum integrated photonics. We consider $Si$ on $SiO_{2}$ waveguides of height $h=340nm$ and base widths of $w_{0}=220nm$. The resonant coupling $\Omega$ is assessed via a finite difference eigenmode (FDE) solver \cite{Lumerical} by evaluating the symmetric $\beta_{+}$ and anti-symmetric $\beta_{-}$ modes as a function of the distance $g$ between the two identical waveguides, where $\Omega = (\beta_{+}-\beta_{-})/2$ (see Fig. \ref{Figure:SFig2}(I)). 

In order to incorporate a phase-mismatch, we increase or decrease the base width $w = w_{0} \pm \delta$ to create a positive or negative detuning value. The propagation constant for each individual waveguide $\beta_{i}$ is calculated via a FDE solver for each width $w_{i}$ to evaluate the resulting detuning $\Delta = (\beta_{1}-\beta_{2})/2$ (see Fig. \ref{Figure:SFig2}(II). The detuning values $\Delta$ normalized by the relevant coupling $\kappa$ are used to asses the relevant differences in width $\delta$ that are required for the detuning modulation. The implementation of detuning-modulated CPs on coupled waveguides is based on the coupled mode equations, therefore the values $\delta$ achieved here are only an approximation to those that comply with the actual light propagation. Therefore, $\delta$ is tweaked via an iterative eigenmode expansion (EME) simulation process to achieve the correct values. Since detuning-modulated CPs are robust to systematic errors, which include target detuning values, the iterative process is short and straightforward.

\end{document}